\documentclass[aps,prl,twocolumn,superscriptaddress]{revtex4-2}

\usepackage{graphicx}
\usepackage{amsmath}
\usepackage{amssymb}
\usepackage{color}
\usepackage{natbib}
\usepackage{siunitx}
\usepackage{comment}
\usepackage{float}

\usepackage[version=4]{mhchem}

\begin{document}

\title{An Ice Christmas Tree: Fast Three-Dimensional Printing of Ice Structures via Evaporative Cooling in Vacuum}

\newcommand{\equaldagger}{\textsuperscript{\textdagger}}

\author{Menno Demmenie\mbox{\equaldagger}}
\author{Stefan Kooij\mbox{\equaldagger}}
\author{Daniel Bonn}

\affiliation{Institute of Physics, University of Amsterdam,
Science Park 904, 1098 XH Amsterdam, Netherlands\\[3pt]
\textsuperscript{\textdagger}\,These authors contributed equally to this work.}

\renewcommand{\andname}{\unskip, } 

\date{\today}

\begin{abstract}
We demonstrate a novel approach to three-dimensional (3D) printing of freeform ice structures by exploiting evaporative cooling. A micrometer-sized water jet is used to 3D print inside a vacuum chamber. The reduced ambient pressure leads to rapid evaporation of the extruded water, extracting latent heat, and quickly cooling the water well below 0 °C. Once deposited, the water freezes almost instantaneously into stable ice structures. We demonstrate high-fidelity printing of complex geometries (Christmas trees, cones, vertical pillars, and free-standing zigzag structures) without cryogenic infrastructure, supporting materials, or external refrigeration. This approach directly visualizes fundamental thermodynamic principles---latent heat, evaporative cooling, and pressure-dependent phase transitions---while offering a relatively simple and scalable platform for ice-templated microfluidics and tissue engineering, or even extraterrestrial 3D printing.
\end{abstract}

\maketitle

Three-dimensional (3D) printing of ice has attracted recent interest as a route to complex, intricate scaffolds for tissue engineering, microfluidic devices, and casting templates \cite{zheng2020inkjet,garg2022freeform,kamble2022multi,Deville2006ice,li2022recent}. Conventional approaches rely on cryogenic refrigeration (using liquid nitrogen or helium) or substrate cooling to stabilize solid water, which requires specialized equipment and infrastructure, and often entails significant operational costs. Here, we introduce a fundamentally different technique by exploiting the rapid evaporative cooling of water under reduced pressure to enable spontaneous freezing of water deposited by a micron-sized jet, requiring only a vacuum pump and a modified commercial 3D printer. 
Importantly, only a very small fraction of the water needs to evaporate to induce freezing---the printed volume does not visibly diminish before solidification, nor does the ice disappear quickly once formed. 
\begin{figure}[h!]
  \centering
  \includegraphics[width=0.8\columnwidth, trim=10cm 4.5cm 8cm 9cm, clip]{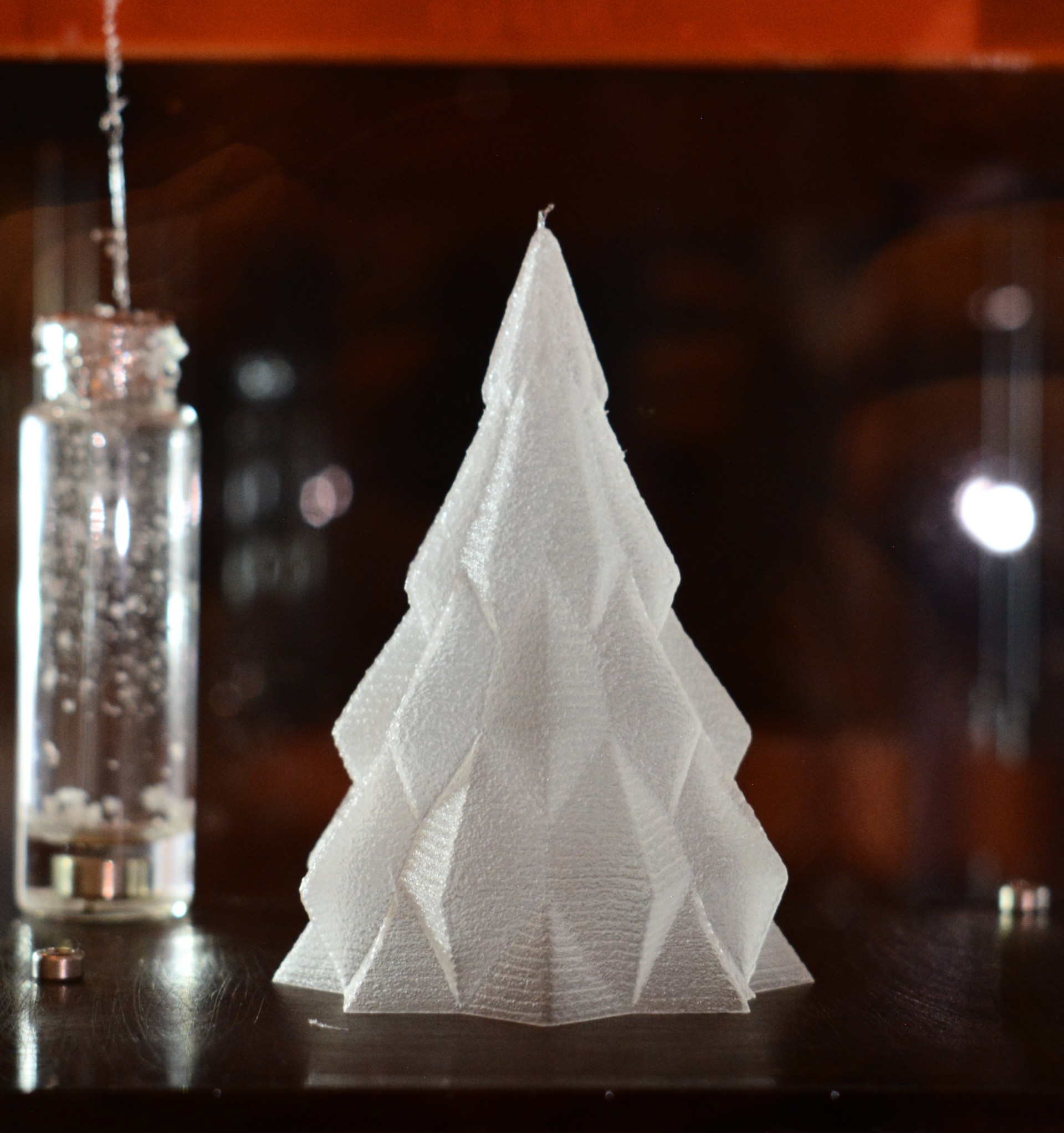}
  \caption{
 Photograph of a Christmas tree made entirely from 3D-printed ice. The structure is fabricated using a 16~\textmu m liquid jet mounted on a commercial 3D printer inside a vacuum chamber. The tree has a height of approximately 8~cm and a base diameter of about 6~cm, with branches and fine features faithfully reproduced from the digital model without any supporting material or external cooling. The slight translucency and smooth surfaces demonstrate the optical quality of the ice and the stability of the evaporative-cooling printing process.
  }
  \label{fig:icetree}
\end{figure}
\\
\indent Interestingly, the low ambient pressures on the Moon ($10^{-12}-10^{-13}$ \SI{}{\milli\bar}) and Mars (\SI{6}{\milli\bar}) are within the range required for this freezing mechanism \cite{stern1999lunar,warren2019through}. This suggests that water-based or particle-laden jets could be used to directly print structures in these environments without additional cooling systems \cite{liu2022effect}. In particular, ice mixed with sand or regolith---similar to natural permafrost---can form a strong composite material \cite{vasiliev2015review,lange1983dynamic}, potentially useful for building protective structures or radiation shielding using local resources, especially when 3D infill geometries are optimized to minimize water use.\\
\begin{figure*}[t!]
    \centering
    \includegraphics[width=0.99\linewidth]{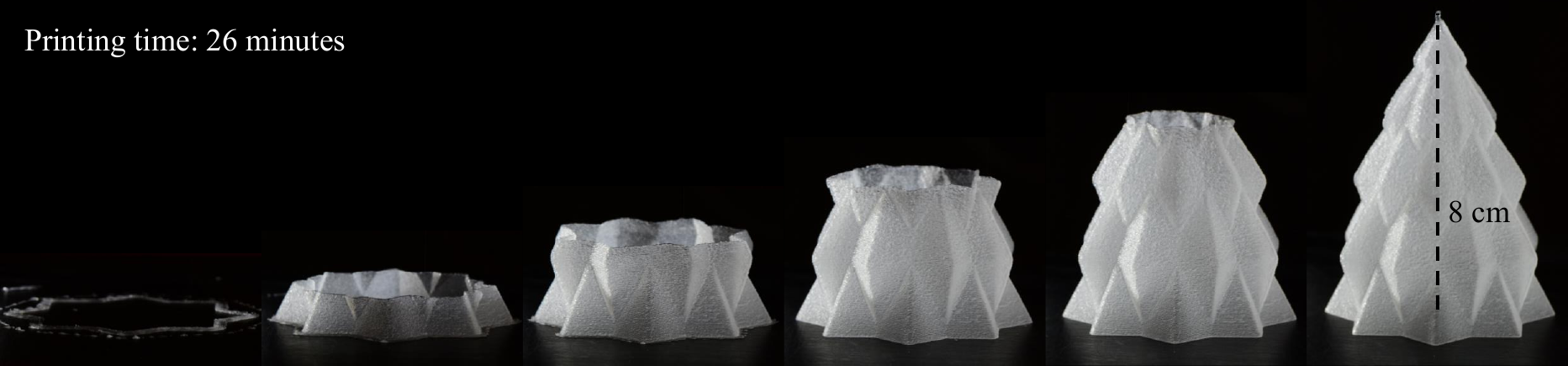}
    \caption{Time-lapse of the 3D-printing process leading up to an ice Christmas, similarly to the tree shown in Figure 1. Total construction time was 26 minutes at 2 mbar.}
    \label{fig:ctree-timelapse}
\end{figure*}
\indent Figure ~\ref{fig:icetree} shows the primary results. A Christmas tree structure was printed in spiral mode to a height of approximately 8 cm, with a base diameter of 6 cm and fine detail features (branches, asymmetries) reproduced from the 3D model. The method is remarkably simple: it requires no external cooling coils, no cryogenic gas infrastructure, and no supporting sacrificial material to stabilize the structure. 
Figure~\ref{fig:ctree-timelapse} shows a time-lapse sequence of the printing process; the complete tree required 26 minutes of printing time. The full printing process is captured in the supplementary movie \texttt{treeprinting.mp4}, while its melting sequence when the vacuum is removed is provided in \texttt{treemelting.mp4}.

\begin{figure}[b!]
  \centering
  \includegraphics[width=0.95\columnwidth]{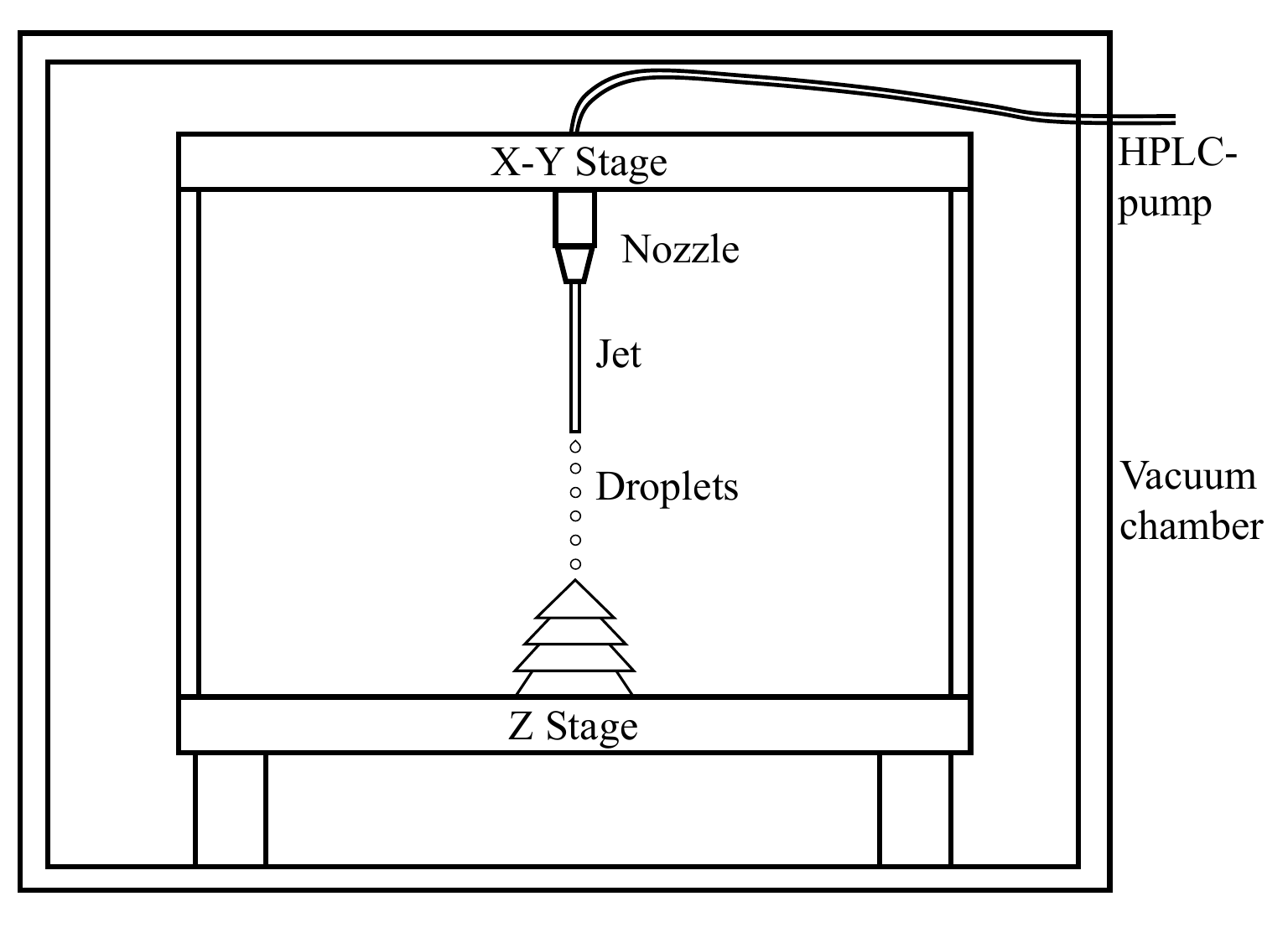}
  \caption{Schematic representation of the 3D ice printing setup. A vacuum chamber houses a 3D positioning system where a nozzle, driven by an High-Performance Liquid Chromatography (HPLC) pump, ejects a 16 µm wide water jet. The jet breaks into droplets that freeze onto the substrate via evaporative cooling to form a solid structure.}
  \label{fig:setup}
\end{figure}

Our setup consists of a commercial 3D printer (ROOK MK1, motion control via open-source firmware) housed inside a transparent acrylic vacuum chamber (Sanatron, inner dimensions: 0.38 m $\times$ 0.36 m $\times$ 0.37 m), as shown in Fig.~\ref{fig:setup}. The standard thermoplastic extruder of the 3D printer was replaced with a custom nozzle mount designed to hold a standard PEK Luer-lock HPLC-adapter. Water is supplied by an HPLC pump, connected to the nozzle via capillary tubing. A \SI{0.45}{\micro\meter} syringe filter can be used to dampen pressure fluctuations during the pump cycles, which otherwise manifest as periodically appearing thicker layer lines in the print (see Fig.~\ref{fig:icetree} as an example without filter). The nozzle consists of a  microfabricated \ce{Si3N4} chip with a \SI{16}{\micro\meter} diameter orifice (Medspray), produced by photolithography and mounted on a polypropylene adapter, which generates a stable laminar water jet with a typical velocity of $v \sim 8$ m/s. The vacuum chamber is evacuated using a rotary vane pump (Oerlikon Leybold SOGEVAC SV 16 D) and the chamber pressure is monitored continuously using an analogue manometer and kept at around 2-3 mbar during printing runs. The heated bed was replaced by a laser-cut plastic plate, the surface of which was lightly sanded to facilitate ice nucleation and crystallization. The print bed is not actively cooled, as the evaporative cooling is sufficient to maintain the print object at freezing temperature. However, a pre-cooled or actively cooled print bed could improve adhesion of the printed object to the bed. Distilled water containing a small amount of NaCl ($>$\SI{1e-4}{M}) was used, since purely distilled water could lead to self-charging, which would negatively affect droplet deposition at a non conductive environment \cite{kooij2022self}.

\begin{figure}[b!]
  \centering
  \includegraphics[width=0.8\columnwidth, trim=8cm 16cm 13cm 12cm, clip]{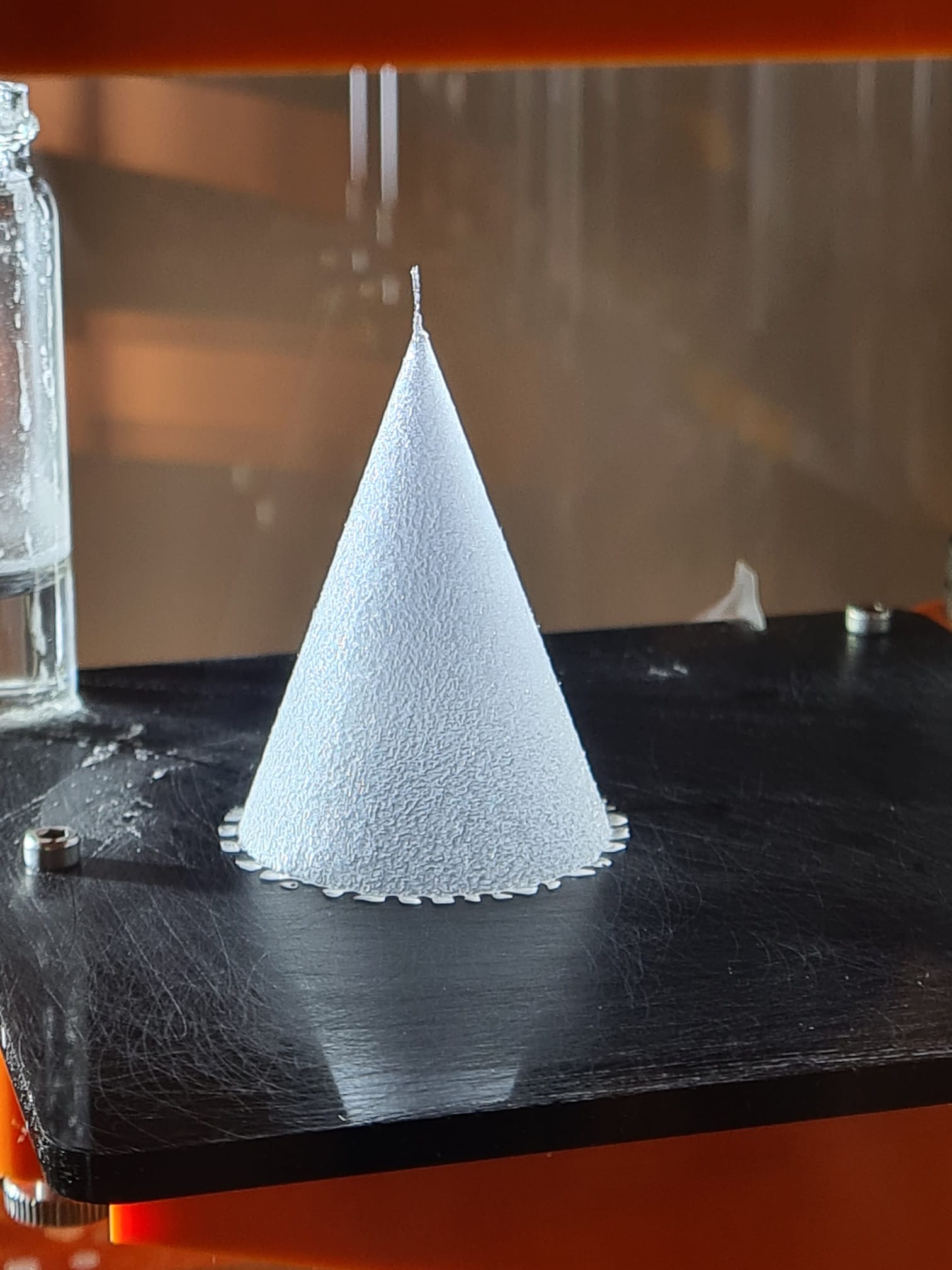}
  \caption{
Photograph of a conical ice structure printed with the vacuum-based 3D ice printer.
  The cone (height $\sim 8$ cm, base diameter $\sim 5$ cm) demonstrates that the method can reproducibly
  generate smooth, axisymmetric geometries without supporting material, with layer lines below the optical
  resolution of the image and a uniformly translucent ice surface.
  }
  \label{fig:icecone}
\end{figure}

A second geometry, a simple cone (base diameter $\approx$ 5 cm, height $\approx$ 8 cm), was printed to demonstrate shape versatility (Figure ~\ref{fig:icecone}). The cone exhibits smooth tapered surfaces with minimal layer-to-layer irregularities, indicating good nozzle trajectory control and stable jet deposition.

In order to 3D print ice without additional cooling, the absorption of latent heat during evaporation must be sufficient to cool the remaining liquid to the desired temperature below 0°C. This energy balance is governed by: 
\begin{equation}
m_{\text{evap}} \cdot c_p \frac{dT}{dt} = - L_v \frac{dm_{\text{evap}}}{dt},
\end{equation}
where $m_{\text{evap}}$ is the evaporated mass, $\rho = 1000$ kg/m$^3$ is the density, $c_p \approx 4.2$ kJ/(kg$\cdot$K) is the specific heat capacity, $L_v = 2.45$ MJ/kg is the latent heat of vaporization, and $dm_{\text{evap}}/dt$ is the mass evaporation rate. For a jet of radius $r = 8$ \textmu m and surface area $A \approx 2\pi r L$ (where $L$ is travel distance), evaporation of less than 5\% of the jet mass per centimeter of travel can already account for the observed temperature drop.

\begin{figure}[t!]
    \centering
    \includegraphics[width=0.95\linewidth]{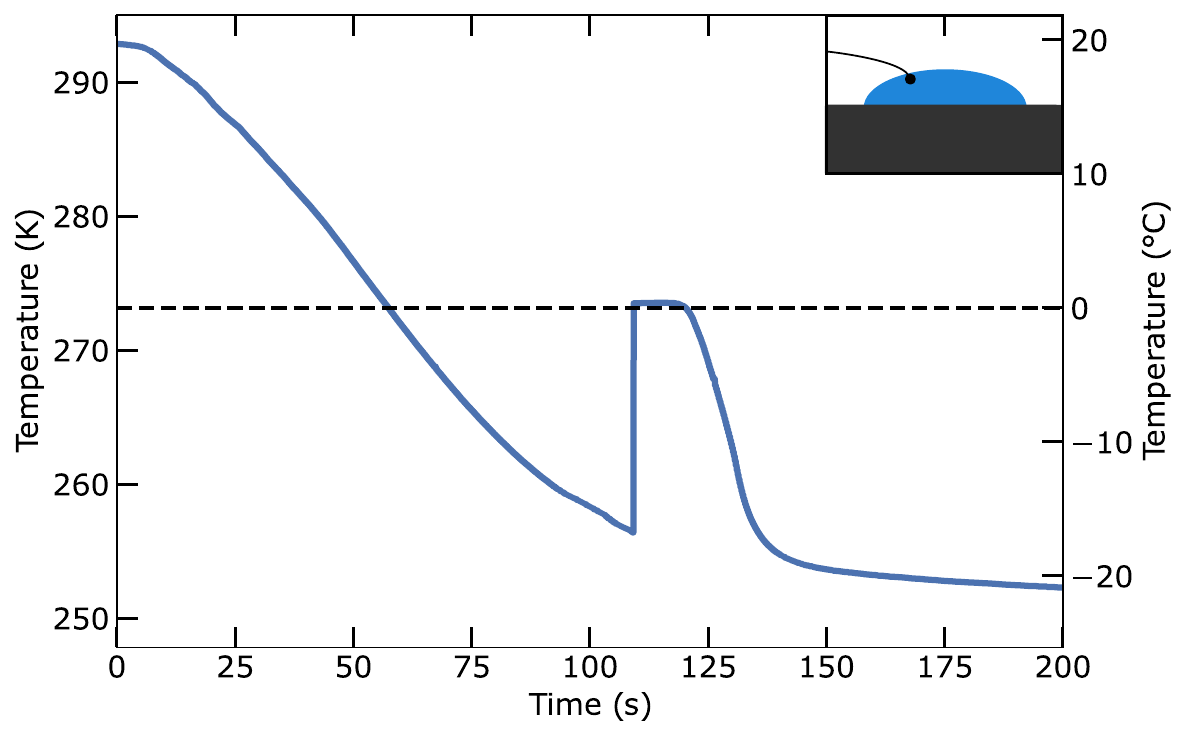}
    \caption{Evaporative cooling of a sessile drop under similar vacuum conditions (2-3 mbar) as used for 3D printing. A macroscopic droplet of 3 mm in diameter was chosen to enable contact thermometry (Pico TC-08), serving as a proxy for the smaller 3D-printing droplets. Recalescence occurs at 110 seconds; the droplet freezes rapidly from an undercooled state, releasing latent heat that drives a sudden temperature rise to $0^{\circ}\text{C}$ (dotted line). The inset shows the experimental schematic.}
    \label{fig:sessile-drop-latentheat}
\end{figure}

To understand this process, we perform a simple reference experiment in which we freeze a mm-sized drop in the same vacuum setup. A simple scaling argument shows that the observed freezing distance of the 16~\textmu m jet is fully consistent with the measured cooling of a millimetric drop in the same vacuum environment. We performed contact thermometry experiments on a sessile water drop of radius $(R = 3~\text{mm}\)  (Fig.~\ref{fig:sessile-drop-latentheat})) which reveals evaporative cooling from room temperature to zero degrees over a timescale of \(\approx 50~\text{s}\), corresponding to a characteristic cooling rate \(\dot{T}_{\mathrm{drop}} \approx 0.4~\text{K/s}\). For our spherical cap of 3 mm radius with a contact angle of $50 ^oC$ , the surface-to-volume ratio which sets the evaporation rate is  $\approx 1600 ~\text{m}^{-1}$. For the liquid jet of 5 cm length and jet diameter \(16~\text{\textmu m}\) the surface to volume ratio is about a factor of 2500 larger than for the millimetric drop. If the evaporative cooling rate scales linearly with the surface to volume ratio, the jet therefore requires only 0.3 s. This estimate is conservative, as it neglects the difference between the way the drop and jet experiments are performed: the drop starts from atmospheric pressure, and is subsequently pumped to $\approx$ 3 mbar, which takes on the order of the time necessary for the drop to freeze; the jest is directly made at 3 mbar and therefore evaporates faster.

\begin{figure*}[]
    \centering
    \includegraphics[width=0.9\linewidth]{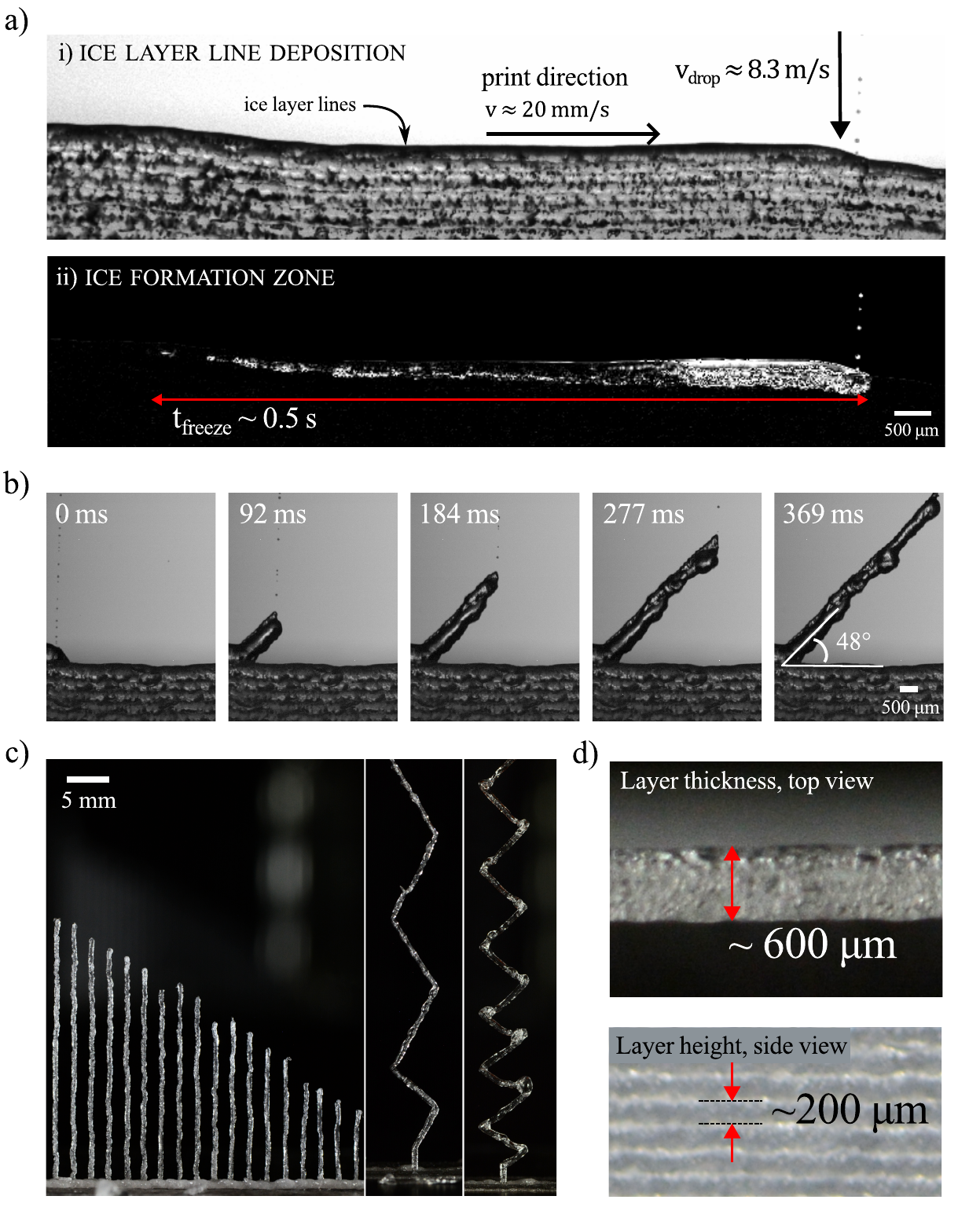}
    \caption{a) High-speed imaging snapshot showing the deposition of micro droplets onto previously printed ice layer lines (side view). The droplet impact velocity ($\approx$\SI{8.3}{\meter\per\second}) is significantly higher than the printer’s travel speed ($\approx$\SI{20}{\milli\meter\per\second}).The upper subpanel (i) shows the raw high-speed image, while the lower subpanel (ii) shows the same frame after subtraction of an earlier image, revealing the ice formation zone. This indicates that the newly deposited line takes approximately \SI{0.5}{\second} to fully freeze. b) When the printer’s travel speed is reduced, free-standing ice lines can be printed. c) By keeping the printer nozzle stationary at specific positions, ice pillars of varying height can be grown. Slowly moving the nozzle during this process produces zigzag-shaped pillars. d) The ice layer height and wall thickness depend on both the print speed and the jet flow rate. For one representative setting, the wall thickness was approximately \SI{600}{\micro\meter} (top view, upper panel) and the layer height \SI{200}{\micro\meter} (side view, lower panel).}
    \label{fig:highspeed_footage}
\end{figure*}

In agreement with these estimates, high-speed imaging at 110,000 fps (Fig.~\ref{fig:highspeed_footage}a) indeed reveals that drops that form from the jet indeed arrive as liquid water on the printed structure, and then freeze completely in about 0.5 s. The high-speed imaging shows that $v_{\textrm{drop}} \approx$ 8.3 m/s, which 400 times faster than the lateral 3D-printing speed. This difference leads to an accumulation of multiple droplets of water on top of the previously printed ice layer. Consequently, the structural accuracy of the print relies heavily on the partial wetting behavior of liquid water on ice. As established in recent experimental studies, water maintains a finite macroscopic contact angle on ice rather than spreading into a thin film \cite{demmenie2023growth, thievenaz2020retraction,demmenie2025partial,sarlin2025macroscopic}. A direct observation of this freezing process is provided by image subtraction, which visualizes the ice formation zone in the lower panel of Fig.~\ref{fig:highspeed_footage}a. The subtraction suppresses static features (shown in black) and highlights regions where change occurs. As indicated by the red arrow, this liquid zone spans approximately 10 mm.

Demonstrating the potential for complex printing architectures, Fig.~\ref{fig:highspeed_footage}b shows the direct fabrication of a complex geometry. A sequence of high-speed images captures the growth of a single ice strut, extending from the base at an angle of 48$^\circ$. This deviation from the horizontal plane is achieved by reducing the printing speed, which enables the water to accumulate and freeze upwards. Notably, this technique enables the direct fabrication of such stable overhangs without auxiliary support structures, overcoming a common limitation in conventional additive manufacturing. Fig.~\ref{fig:highspeed_footage}c furthermore illustrates the geometric versatility of the 3D-printing technique through a series of high-aspect-ratio features ranging from vertical poles to intricate zigzag formations. These examples confirm that the printed ice maintains sufficient mechanical strength to support its own weight. Moreover, these structures demonstrated remarkable flexibility: when the printer was shaken, the slender ice struts oscillated with large amplitudes without breaking. The ability to realize such delicate, unsupported geometries opens the door to creating complex negative molds. For example, printed spiral or branching ice structures can be used as sacrificial templates: after casting a material around them, the ice is simply melted away to reveal hollow channels, such as artificial arteries or microfluidic devices.

To evaluate the geometric consistency of the printing process, the dimensions of the ice walls were analyzed. In Fig.~\ref{fig:highspeed_footage}d, the top view shows a continuous printed track width of $\sim$600 \textmu m. In the side view, the individual layers are clearly distinguishable, indicating that the freezing kinetics allowed for shape retention. The measured layer height of $\sim$200 \textmu m corresponds to the vertical resolution of the current printing settings. Note that for this first demonstration, the nozzle size was chosen for robustness and simplicity; employing a printing system with smaller nozzles, higher mechanical stability and faster translation speeds should enable the fabrication of even thinner structures.

In summary, we have demonstrated a 3D ice-printing method that differs from existing techniques, which typically rely on cooled substrates or subzero ambient temperatures, both involving substantial cooling infrastructure \cite{li2022recent}. Our vacuum evaporative approach requires no dedicated cryogenic infrastructure. Instead it relies on readily available vacuum and 3D-printing components, thus keeping operating costs well below those of liquid-nitrogen- or helium-based systems. The method exploits the thermodynamic instability of liquid water at low pressure, where rapid evaporation extracts latent heat and cools the extruded water far below 0°C. The printed structures consist of pure ice, containing no supporting material, dopants, or foreign particles, which is crucial for applications in microfluidics and tissue engineering where material purity directly impacts biocompatibility and fluid properties. Interestingly, the thin printed ice structures exhibit remarkable flexibility: long, slender struts were observed to vibrate with large amplitudes without breaking. This suggests that tailored geometries could be used to realize ice structures with mechanical responses very different from those of bulk ice, opening opportunities for both functional and aesthetic applications. 

The method is also intrinsically scalable: independent control over jet diameter, flow rate, chamber pressure, and printing speed allows both resolution and production rate to be tuned as needed. Finally, the output is exceptionally clean; once the chamber is vented, the printed ice melts back to liquid water without leaving any residue, making it straightforward to integrate the process with downstream fabrication or processing steps.

From a materials perspective, ice-templated structures have long been used to create hierarchical porous materials \cite{Deville2006ice,mukai2004formation}. Freeze-casting, where suspensions are directionally frozen and then dried, produces anisotropic porous ceramics and composites with tailored pore structure and mechanical properties. Our method offers a complementary approach: instead of freezing a suspension in a mold, we sculpt pure ice directly, which can then serve as a sacrificial template for microfluidic channel networks \cite{Whitesides2006microfluidics}, tissue scaffolds \cite{Langer1993tissue}, or complex casting forms. 


Finally, the underlying physics is relevant to atmospheric science and ice nucleation. Ice formation in clouds occurs via nucleation of water droplets in the upper troposphere, where pressures and temperatures are low. Our vacuum evaporation cooling demonstrates a complementary ice formation pathway: at moderate supersaturation over water (achievable in the upper atmosphere), even pure liquid droplets can spontaneously freeze if the cooling is rapid enough \cite{Koop2000homogeneous}. Our macroscopic demonstration provides a conceptual model for understanding nucleation at microscale.
\\

\noindent Several extensions of this work are foreseeable:

\textit{Resolution enhancement:}
As a standard 3D printer was used, the printing accuracy is primarily limited by mechanical factors such as backlash, vibrations, and dimensional tolerances. Moreover, the jet velocity is much higher than the printer’s travel speed, causing droplets to accumulate at their landing position while the print head moves. This results in walls that are significantly thicker than the jet diameter, with the final wall thickness mainly determined by the wetting properties of water on ice. Although smaller jets could easily be implemented, our current system lacks the positional accuracy required to take advantage of them. A more advanced motion system, capable of high-speed and highly precise positioning, could enable an entirely different printing regime—true droplet-by-droplet deposition—potentially facilitated by stimulated jet breakup to produce a monodisperse droplet train. These higher printing resolutions could reach sub-millimeter scales, enabling intricate micro-architected ice structures for microfluidic and photonic applications.

\textit{Bridging:}
For many 3D printing techniques, bridging presents a major challenge in fabricating complex structures. Here, we demonstrate that, beyond conventional layer-by-layer deposition, curved, angled, and straight ice pillars can be formed simply by reducing the print velocity while maintaining the same jet velocity. This principle can be exploited to create support elements or bridged structures, and greater control over this mechanism could prove invaluable for future applications.

\textit{Hybrid printing:} Water could be mixed with dissolved polymers, salts, or nanoparticles before injection, allowing the deposition of composite ice--material structures. Upon melting or selective sublimation, these could yield polymer or inorganic scaffolds with controlled geometry.

\textit{Layer line smoothing:}
Layer-line smoothing in FDM printing is commonly used to improve the visual appearance of printed objects. For ice, this process would be particularly simple: partially melting and refreezing the surface of the printed object would result in a smoothened ice print.

\textit{Planetary applications:} The Martian polar regions contain water ice and operate under near-vacuum conditions (pressures $\sim$1--10 mbar, temperatures $\sim -100$ to $-150$°C). Our method could be adapted for in situ resource utilization (ISRU), using Martian water to construct habitats, thermal shielding, or radiation barriers without importing external materials or refrigeration systems.

\textit{Tissue engineering and regenerative medicine:} Pure ice scaffolds with controlled pore size and interconnectivity can be used as temporary supports for cell seeding, vascularization, and tissue growth. Upon melting, the scaffold dissolves without chemical residue, leaving behind organized tissue.

In sum, we have demonstrated a simple, elegant, and cryogen-free method for three-dimensional printing of pure ice structures via evaporative cooling in vacuum. A commercial 3D printer, modified with a fine water jet nozzle fed by an HPLC pump and housed in a acrylic vacuum chamber, produces complex ice geometries (Christmas tree, cone, and other designs) with high fidelity. The underlying physics rests on the thermodynamic instability of liquid water at low pressure, where rapid evaporation extracts latent heat and cools the extruded water far below 0 °C. 

The method requires no cryogenic infrastructure, no supporting material, and no external refrigeration, making it accessible to many laboratories and enabling new applications in microfluidics, tissue engineering, and materials templating. The visual demonstration of evaporative cooling and phase transitions also serves as a powerful educational tool.


This work opens a new frontier in additive manufacturing of ice and demonstrates the power of exploiting simple thermodynamic principles for engineering innovation. Future directions include miniaturization for high-resolution microstructures, hybrid printing with dissolved additives, and adaptation to planetary environments such as found on Mars. 
\\
\subsection{Acknowledgment}
The authors would like to thank Maarten Tuijl for technical support, Christiaan van Campenhout for assisting with visuals, and Medspray for providing customized nozzles. This project has received funding from the European Research Council (ERC) under the European Union's Horizon 2020 research and innovation programme (Grant agreement No. 101142159).

\bibliographystyle{aipnum4-1}
\bibliography{biblio}

@article{kooij2022self,
  title={Self-charging of sprays},
  author={Kooij, Stefan and van Rijn, Cees and Ribe, Neil and Bonn, Daniel},
  journal={Scientific Reports},
  volume={12},
  number={1},
  pages={19296},
  year={2022},
  publisher={Nature Publishing Group UK London}
}

@article{lange1983dynamic,
  title={The dynamic tensile strength of ice and ice-silicate mixtures},
  author={Lange, Manfred A and Ahrens, Thomas J},
  journal={Journal of Geophysical Research: Solid Earth},
  volume={88},
  number={B2},
  pages={1197--1208},
  year={1983},
  publisher={Wiley Online Library}
}

@article{stern1999lunar,
  title={The lunar atmosphere: History, status, current problems, and context},
  author={Stern, S Alan},
  journal={Reviews of Geophysics},
  volume={37},
  number={4},
  pages={453--491},
  year={1999},
  publisher={Wiley Online Library}
}

@article{warren2019through,
  title={Through the thick and thin: New constraints on Mars paleopressure history 3.8--4 Ga from small exhumed craters},
  author={Warren, AO and Kite, ES and Williams, J-P and Horgan, B},
  journal={Journal of Geophysical Research: Planets},
  volume={124},
  number={11},
  pages={2793--2818},
  year={2019},
  publisher={Wiley Online Library}
}

@article{liu2022effect,
  title={Effect of the freezing temperature and near-vacuum air pressure of Mars on the mechanical properties and microstructure of hydrogel-based concrete (HBC)},
  author={Liu, Ning and Qiu, Jishen},
  journal={Extreme Mechanics Letters},
  volume={56},
  pages={101864},
  year={2022},
  publisher={Elsevier}
}

@article{Deville2006ice,
  title={Ice-templated materials},
  author={Deville, S. and Saiz, E. and Nalla, R. K. and Tomsia, A. P.},
  journal={Science},
  volume={311},
  number={5760},
  pages={515--518},
  year={2006},
  publisher={American Association for the Advancement of Science},
  doi={10.1126/science.1124237}
}

@article{Koop2000homogeneous,
  title={Water activity as the determinant of homogeneous ice nucleation in aqueous solutions},
  author={Koop, T. and Luo, B. and Tsias, A. and Peter, T.},
  journal={Nature},
  volume={406},
  number={6796},
  pages={611--614},
  year={2000},
  doi={10.1038/35020537}
}

@article{Whitesides2006microfluidics,
  title={The origins and the future of microfluidics},
  author={Whitesides, G. M.},
  journal={Nature},
  volume={442},
  number={7101},
  pages={368--373},
  year={2006},
  publisher={Nature Publishing Group},
  doi={10.1038/nature05058}
}

@article{Langer1993tissue,
  title={Tissue engineering},
  author={Langer, R. and Vacanti, J. P.},
  journal={Science},
  volume={260},
  number={5110},
  pages={920--926},
  year={1993},
  publisher={American Association for the Advancement of Science},
  doi={10.1126/science.260.5110.920}
}

@article{garg2022freeform,
  title={Freeform 3D Ice Printing (3D-ICE) at the Micro Scale},
  author={Garg, Akash and Yerneni, Saigopalakrishna S and Campbell, Phil and LeDuc, Philip R and Ozdoganlar, O Burak},
  journal={Advanced Science},
  volume={9},
  number={27},
  pages={2201566},
  year={2022},
  publisher={Wiley Online Library}
}

@article{kamble2022multi,
  title={Multi-jet ice 3D printing},
  author={Kamble, Pushkar Prakash and Chavan, Subodh and Hodgir, Rajendra and Gote, Gopal and Karunakaran, KP},
  journal={Rapid Prototyping Journal},
  volume={28},
  number={6},
  pages={989--1004},
  year={2022},
  publisher={Emerald Publishing Limited}
}

@article{zheng2020inkjet,
  title={Inkjet printing-based fabrication of microscale 3D ice structures},
  author={Zheng, Fengyi and Wang, Zhongyan and Huang, Jiasheng and Li, Zhihong},
  journal={Microsystems \& Nanoengineering},
  volume={6},
  number={1},
  pages={89},
  year={2020},
  publisher={Nature Publishing Group UK London}
}

@article{vasiliev2015review,
  title={A review on the development of reinforced ice for use as a building material in cold regions},
  author={Vasiliev, NK and Pronk, ADC and Shatalina, IN and Janssen, FHME and Houben, RWG},
  journal={Cold Regions Science and Technology},
  volume={115},
  pages={56--63},
  year={2015},
  publisher={Elsevier}
}

@article{sarlin2025macroscopic,
  title={The macroscopic contact angle of water on ice},
  author={Sarlin, Wladimir and Papa, Daniel Vito and Grivet, Rodolphe and Rosenbaum, Alexander and Huerre, Axel and S{\'e}on, Thomas and Josserand, Christophe},
  journal={Journal of Fluid Mechanics},
  volume={1019},
  pages={A31},
  year={2025},
  publisher={Cambridge University Press}
}

@article{demmenie2025partial,
  title={Partial wetting of water on ice},
  author={Demmenie, Menno and Gorin, Benjamin and Kolpakov, Paul and Smith, Scott and Kellay, Hamid and Bonn, Daniel},
  journal={Physical Review Fluids},
  volume={10},
  number={5},
  pages={054002},
  year={2025},
  publisher={APS}
}

@article{demmenie2023growth,
  title={Growth and form of rippled icicles},
  author={Demmenie, Menno and Reus, Lars and Kolpakov, Paul and Woutersen, Sander and Bonn, Daniel and Shahidzadeh, Noushine},
  journal={Physical Review Applied},
  volume={19},
  number={2},
  pages={024005},
  year={2023},
  publisher={APS}
}

@article{thievenaz2020retraction,
  title={Retraction and freezing of a water film on ice},
  author={Thi{\'e}venaz, Virgile and Josserand, Christophe and S{\'e}on, Thomas},
  journal={Physical Review Fluids},
  volume={5},
  number={4},
  pages={041601},
  year={2020},
  publisher={APS}
}

@article{li2022recent,
  title={Recent advances in cryogenic 3D printing technologies},
  author={Li, Zongan and Xu, Mengjia and Wang, Jiahang and Zhang, Feng},
  journal={Advanced Engineering Materials},
  volume={24},
  number={10},
  pages={2200245},
  year={2022},
  publisher={Wiley Online Library}
}

@article{mukai2004formation,
  title={Formation of monolithic silica gel microhoneycombs (SMHs) using pseudosteady state growth of microstructural ice crystals},
  author={Mukai, Shin R and Nishihara, Hirotomo and Tamon, Hajime},
  journal={Chemical communications},
  number={7},
  pages={874--875},
  year={2004},
  publisher={Royal Society of Chemistry}
}

\end{document}